\documentclass[amsmath,amssymb,amsfonts,aps,pre,superscriptaddress,bibnotes,showpacs,showkeys,longbibliography,10pt,twocolumn]{revtex4-2}
\usepackage[utf8]{inputenc}
\usepackage[T1]{fontenc}
\usepackage{caption,dsfont}
\usepackage{mathtools}
\usepackage{epsfig}
\usepackage[english]{babel}
\usepackage{color}
\usepackage{bbold}
\usepackage{subcaption}
\usepackage{lipsum}
\usepackage{array,multirow}
\usepackage{booktabs}
\usepackage{tikz}
\usetikzlibrary{braids}
\usepackage{amsthm}
\usepackage{physics}
\usepackage{float}

\usepackage{epstopdf}
\usepackage[colorlinks,linkcolor=blue,citecolor=blue]{hyperref}  \usepackage{url}

\AppendGraphicsExtensions{.tga}

\graphicspath{{Figures/}}

\newcommand{\wt}{\mathrm{W}}
\newcommand{\es}{\mathcal{O}}
\newcolumntype{M}[1]{>{\centering\arraybackslash}m{#1}}

\begin{document}

\title{Aggregation and structural phase transitions of semiflexible polymer bundles: a braided circuit topology approach}

\author{Jonas Berx}

\affiliation{Medical Systems Biophysics and Bioengineering, Leiden Academic Centre for Drug Research, Faculty of Science, Leiden University, 2333CC Leiden, The Netherlands}
\author{Alireza Mashaghi}%
\thanks{Lead Contact}
\email{a.mashaghi.tabari@lacdr.leidenuniv.nl}
\affiliation{Medical Systems Biophysics and Bioengineering, Leiden Academic Centre for Drug Research, Leiden University, 2333CC Leiden, The Netherlands}

\begin{abstract}
We present a braided circuit topology framework for investigating topology and structural phase transitions in aggregates of semiflexible polymers. In the conventional approach to circuit topology, which specifically applies to single isolated folded linear chains, the number and arrangement of contacts within the circuitry of a folded chain give rise to increasingly complex fold topologies. Another avenue for achieving complexity is through the interaction and entanglement of two or more folded linear chains. The braided circuit topology approach describes the topology of such multiple-chain systems and offers topological measures such as writhe, complexity, braid length, and isotopy class. This extension of circuit topology to multichains reveals the interplay between collapse, aggregation, and entanglement. In this work, we show that circuit topological motif fractions are ideally suited order parameters to characterise structural phase transitions in entangled systems that can detect structural re-ordering other measures cannot.
\end{abstract}


\maketitle

\thispagestyle{empty}

\section*{Introduction}\label{sec:intro}

It has been shown that, for a single semiflexible chain, the polymer stiffness leads to a structural ``phase diagram'', displaying a multitude of of conformations such as globular, hairpin, knotted or extended structures, both for off-lattice \cite{seaton2013,polym8090333,Marenz2016} and on-lattice systems \cite{Doniach1996,Bastolla1997}. By keeping a constant monomer-monomer interaction energy scale and decreasing the temperature, the polymer minimizes its energy by either collapsing or stiffening locally. Adjusting the stiffness $\kappa$ allows one to study the competition between these two effects, which leads to fundamentally distinct structural motifs.

This line of reasoning can be extended to aggregates of semiflexible polymers. It has been shown \cite{Zierenberg_2015,polym8090333} that for small systems consisting of $M = 2,4,8$ polymers of length $N = 13$ at a high temperature $T$, the system is fragmented and individual polymers can be considered isolated. The structural properties in this regime follow the single-chain results in good approximation. For a decreasing temperature, however, flexible polymers aggregate in an amorphous globular configuration, while stiffer polymers form (twisted) bundles. The concomitant disentanglement of long polymers during cooling also depends fundamentally on the stiffness \cite{Luo2012}, the study of which offers insights into effects such as the asymmetry between crystallisation and melting, or glass formation \cite{Nguyen2015}. Moreover, for polymer melts lowering the temperature can lead to a transition from a disordered melt phase to an ordered nematic and crystalline phase, as a consequence of increasing stiffness \cite{Kawak2021,MartnezFernndez2023}.

The collapse and aggregation transitions are not separate processes. If the energy reduction associated with the formation of more interchain contacts (aggregation) is more favourable than the formation of intrachain contacts (collapse or folding), multichain aggregation may be expected to undo single-chain collapse. Hence, by studying the inter-, and intrachain contacts, we can deduce information about the structural properties and phase transitions of the system. Both types of contacts can be neatly studied within the framework of circuit topology (CT), which was originally introduced to describe the topology of folded proteins and folding reactions \cite{MASHAGHI2014,Heidari2020}, and was subsequently extended to include entanglement (i.e., `soft contacts') \cite{GOLOVNEV2020, GolovnevBook2022,Berx2023} and multiple chains \cite{Heidari2022}.
By combining circuit topology with a topological description on the level of entanglement, we get a more complete picture of aggregation processes with a small number of chains.

The outline of this paper is as follows: in section \ref{sec:mCT} we set the stage for our analysis by reiterating the basic concepts of circuit topology for multichain systems. Subsequently, in section \ref{sec:braids}, we discuss the theoretical framework necessary for our concomitant braid-theoretic description. Section \ref{sec:simulation} investigates the structural phase transitions in a system of $M=4$ chains of varying monomer number $N\in\{10,\,30\}$ and connects this with the circuit topology and braiding analysis. Finally, in section \ref{sec:conclusion}, we present conclusions and a future outlook.

\section{Multichain Circuit Topology}\label{sec:mCT}
We start by revisiting multichain circuit topology. To distinguish between the aforementioned `soft contacts', which are units of single-chain entanglement, we will henceforth only consider `hard contacts', which we will hereafter simply refer to as `contacts'. To model the formation of intra- and inter-chain bonds in molecular systems and the resulting topology, we can examine the mutual topological relation of binary contacts that represent these bonds in the circuit topology framework. Two different contacts, that we name $A$ and $B$, each consisting of two contact `sites' bearing the same name - totalling four sites - will be assigned to a topological relation. Note that such `bonds' pertain to any types of atomic or molecular interactions that contribute to folding and assembly, including covalent and non-covalent ones (e.g., hydrogen bonds, disulfide bonds, beta-beta interactions in biomolecules). Depending on the application, one may focus on a specific type of bonds, and model them as binary contacts.

On a single strand, only three possible arrangements of $A$ and $B$ are possible: $AABB$, $ABBA$ and $ABAB$, corresponding respectively to a series ($S$), parallel ($P$) or cross ($X$) motif, where we take renaming  $(A\leftrightarrow B)$ into account.

\begin{figure*}[tp]
    \centering
    \includegraphics[width=\linewidth]{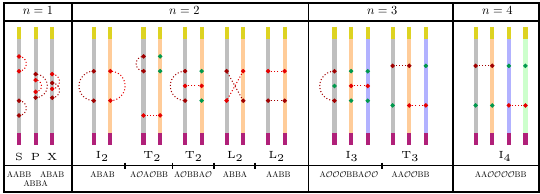}
    \caption{The set of circuit topology motifs for $n=1,2,3,4$ strands. Contacts $A$ and $B$ are indicated by red and turquoise filled circles, respectively, while ghost contacts $\mathcal{O}$ are indicated by filled green circles. Different chains are indicated by different colours. The coloured chain ends indicate the orientation of the strands; we choose to orient every strand from the yellow to the violet end. Corresponding string notations are given below each motif, where the string is read in $n$-tuples.}
    \label{fig:motiflist}
\end{figure*}

Similar to the case of single chains, a motif that comprises pairs of contact sites is referred to as a ``circuit''. A circuit can be braided, as described in subsequent sections. When considering a system consisting of multiple distinct open-ended chains, a number of motifs needs to be added to the above set of circuits. For $n=2$ chains, we can discern three topologically distinct motifs: independent ($I_2$), loop ($L_2$) and tandem ($T_2$). Within these motifs, degeneracies are possible. The independent relation is unique up to renaming of the contact sites, the loop relation can form either a ``parallel'' or a ``cross'' loop (not to be confused with the single-chain CT motifs), and the tandem relation occurs in either an ``umbrella'' or an ``arc'' motif (Fig.~\ref{fig:motiflist}). 

For $n=3$, only independent ($I_3$) and tandem ($T_3$) circuits can be formed, which are both non-degenerate. Lastly, for $n=4$, only the independent relation ($I_4$) can occur. Four is the maximum number of chains that can participate in creating circuits with four contact sites. In Fig.~\ref{fig:motiflist}, we list all possible multichain motifs. 

Extending now the string notation for the single-chain CT to the multichain framework requires the introduction of ``ghost contacts'', which we will denote with $\es$ in the string. To construct the string notation corresponding to a particular motif, we proceed as follows. The two chains are placed parallel and aligned vertically. We now read from top to bottom and from left to right, adding a ghost contact wherever a real contact site does not possess partners on the same level on other chains.

A simple example will clarify our formalism. Consider the $T_2$ umbrella CT motif, i.e., the tandem relation on two chains. We assume that the first chain counted from the left possesses the ordered contact sites $A\,,B\,,A$, and that the second chain only possesses a single $B$ contact site. We supplement the right chain with two ghost nodes. The string describing $T_2$ is then $\mathcal{S}_2 = ABB\es A\es$, where the subscript indicates the number of chains that participate in the motif. Note that the same motif is obtained independent of the exact position of the ghost contacts on the second chain.

The string notations for all multichain motifs can be found in Fig.~\ref{fig:motiflist}. Note that all motifs within a degenerate circuit class (e.g., $T_2$) can be obtained by cyclically permuting the characters corresponding to a single chain.

\section{Topological entanglement and braiding measures \label{sec:braids}}
\begin{figure}[htp]
    \centering
        \includegraphics[width=0.5\linewidth]{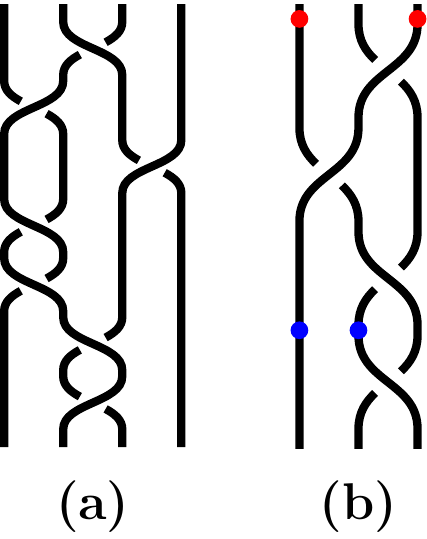}
    \caption{{\bf (a)} A braid with index $n=4$ with string notation $\mathcal{S}_4=\sigma_2\sigma_1^{-1}\sigma_3^{-1}\sigma_1\sigma_1\sigma_2\sigma_2^{-1}$. {\bf (b)} A braid containing contacts $A,\,B$ indicated by red and turquoise points, respectively. The string notation is $\mathcal{S}_3 = A\es A\,\sigma_2^{-1}\sigma_1^{-1}\sigma_2\,BB\es\, \sigma_2$.}
    \label{fig:examplebraid}
\end{figure}

We now assume that the different chains can cross one another in the embedding space, and henceforth call the number of chains the \emph{braid index} $n$, where we will limit ourselves to $n\leq 4$. For $n=2$, a twist of the chains is not topologically protected, i.e., it can be undone by rotating one of the planes in which the endpoints are fixed.  we will henceforth assume that the resulting braid projection is confined to the plane spanned by the parallel lines connecting the first and last contact-contact or contact-ghost pair. We label the chains by $\nu = 1\,,2\,,3\,,...$ from the left and introduce notation that is common in the theory of braiding, by following the Artin representation \cite{BIRMAN2005}. The operators $\sigma_i$, with $i=1,2,...,n-1$, indicate that the $i$th chain, as counted from the left passes \emph{over} the $(i+1)$th chain. The inverse operators $\sigma_1^{-1}\,,\sigma_2^{-1}\,,...\,,\sigma_i^{-1}$ indicate that the $i$th chain passes \emph{underneath} the $(i+1)$th chain. Note that the index $i$ is independent of the labelling $\nu$ of the strands; $i$ represents only the order of the strands, as read from left to right, while $\nu$ represents the strands themselves. This way, we can represent a braid with index $n$ by a string of $\sigma-$operators. For example, the string $\beta = \sigma_2\sigma_1^{-1}\sigma_3^{-1}\sigma_1\sigma_1\sigma_2\sigma_2^{-1}$ corresponds to the braid shown in Fig.~\ref{fig:examplebraid}(a).

We remark here that our string notation for CT and the Artin braid notation can quite naturally be combined into a generic framework by inserting the braid operators $\sigma_i$ in the string after every $n$-tuple of contacts. An example is given in Fig.~\ref{fig:examplebraid}(b). Hard contacts can then be modelled within this framework as separate operators, with their own set of Reidemeister moves and bond moves \cite{Gugumcu2022}. We defer to ongoing theoretical research on this topic and will not use the string notation any further in this manuscript.

Let us formulate some useful properties of braids before continuing. Similar to knot theory, braids belong to the same equivalence class if they are related by a set of moves that are related to the so-called Reidemeister moves. The three fundamental moves can be formulated as follows:
\begin{enumerate}
    \item $\sigma_i\sigma_j = \sigma_j\sigma_i$ for $|i-j|\geq2$ (disjoint strand relation)
    \item $\sigma_i\sigma_{i+1}\sigma_i = \sigma_{i+1}\sigma_i\sigma_{i+1}$ (Skein relation)
    \item $\sigma_i\sigma_i^{-1} = e$ (annihilation relation)
\end{enumerate}
The element $e$ is the unit element of the braid group; if we encounter such an operator pair we can eliminate it from the string. The braid group $B_n$ itself is then defined from these moves. 

A natural, yet nontrivial question now arises: which topological quantities can we compute from this braid representation to characterise the entanglement of our system of semiflexible polymers? We briefly discuss four possible measures that can quantify certain topological properties of a braid: the writhe, braid length, complexity index and isotopy class.

\subsection{Writhe}
The writhe $\wt(\beta)$ of a braid $\beta$ is defined as the sum of the exponents of the braid operators, i.e., for $\beta = \sigma_{i_1}^{a_1}\sigma_{i_2}^{a_2}...\sigma_{i_m}^{a_m}$ it is $\wt(\beta) = \sum_k a_k$. It characterises the net twist of a braid and can be used to detect whether underlying (chemical) chirality of the monomers or interactions between them influences the mesoscopic twist of the system \cite{Lee2013}. When there are no interactions that prefer one handedness over the other, the writhe can be expected to be symmetrically distributed with mean zero. Since two equivalent braids have the same writhe, it is a topological braid invariant. In, e.g., Fig.~\ref{fig:examplebraid}(a) the writhe is $\wt = 1$, while for (b) is is $\wt = 0$.

Since in the next section we will use the Kremer-Grest model, which does not intrinsically include any chiral interactions, to numerically simulate the system, we compute the writhe only as a sanity check.

\subsection{Braid length}
The braid length $L(\beta)$ is a topological measure, not a geometrical one. It counts the number of operators in the braid word $\beta$. Since we can always add the trivial combination $\sigma_i \sigma_i^{-1}$ to the braid word, the length is not an invariant. However, the minimal braid length $L_c(\beta)$, obtained by reducing the braid word to its normal form, is a braid invariant. Henceforth, we will only consider the minimal length and will just refer to it as ``the length $L$'', without ambiguity. In Fig.~\ref{fig:examplebraid}(a), the minimum braid length is $L = 3$, since the normal form of the braid is $\beta_c = \sigma_2\sigma_3^{-1}\sigma_1$.

To some extent, the length can be used as a measure for the complexity of a braid, i.e., the greater the length, the more complex the associated braid. Since it is a feature of the normal form, trivial entanglement such as, e.g., projection-induced crossings $\sigma_i\sigma_i^{-1}$ are eliminated and only actual crossings are counted. Naturally, since we only consider systems with a finite number of monomers, we can assume that the maximum braid length is bounded and that it heavily depends on the number of chains $M$, number of monomers per chain $N$ and the stiffness $\kappa$.

\subsection{Complexity index}
\begin{figure}[htp]
    \centering
    \includegraphics[width=\linewidth]{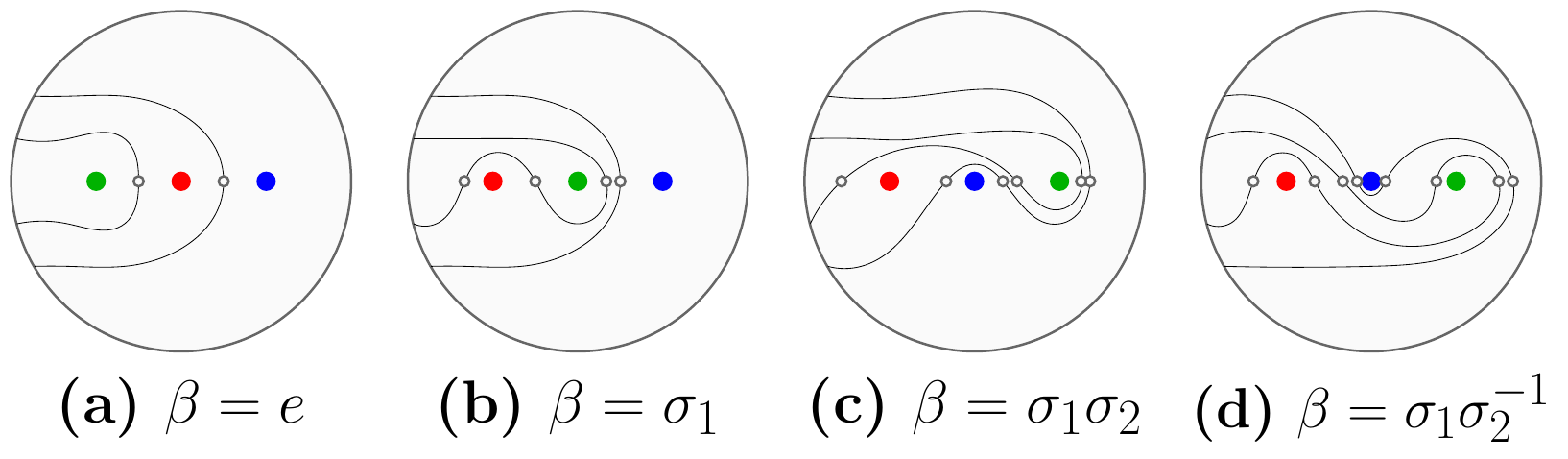}
    \caption{The punctured disk representation of the braid complexity for four different braids. Punctures corresponding to strands are indicated by coloured dots. The operators $\sigma_i,\,\sigma_i^{-1}$ change punctures $i$ and $i+1$ CW or CCW, respectively. Intersections with the central axis are indicated by open circles.}
    \label{fig:punctured_disc}
\end{figure}

Just how entangled is a braid? This is not an easy question to answer, since complexity is not an objective observable. Therefore, we make the choice that henceforth entanglement can be characterised by the complexity index $C(\beta)$ \cite{Dynikov2007}. The braid complexity index, together with the braid length, constitutes our main measure to characterise the topology and entanglement properties of a braid. It is defined as the number of intersections (which we denote by \#) of a curve spanning a punctured disk of $n$ punctures with the real axis, after application of the braiding operations. This is illustrated in Fig.~\ref{fig:punctured_disc}. Normalising by the number of initial intersections $\#E$ with the real axis, which is $n-1 =3$ in our case, and subsequently taking the natural logarithm, the complexity can be written as $C(\beta) = \log\left(\frac{\#\beta E}{\#E}\right)$.

So are for example the braids $\beta_1 = \sigma_1\sigma_2$ and $\beta_2 = \sigma_1\sigma_2^{-1}$ very similar, differing only in the sign of the second crossing, but their complexities are $C_1 = 1.099$ and $C_2 = 1.386$, respectively, as can be seen in Fig.~\ref{fig:punctured_disc}(c)-(d). This indicates that braid $\beta_2$ is more complex (i.e., more entangled) than braid $\beta_1$, which was intuitively clear. In particular, since we only consider systems consisting of $M = 4$ chains, we will use the base-three instead of the natural logarithm, i.e., $C = \log_3(\frac{\#\beta E}{\#E})$. In this formulation, the braid $\beta = \sigma_1\sigma_2^{-1}$ has a complexity equal to unity.

\subsection{Isotopy class}
Every braid can be classified according to three types of isotopy classes given by the Thurston–Nielsen (TN) classification: reducible (RE), finite-order (FO) or pseudo-Anosov (PA) \cite{nielsen1944surface,Thurston1988,Thiffeault2022}. The isotopy class of a braid carries global information about the topology. For instance, if a braid is finite-order or, alternatively, periodic if it can be isotoped to the identity after $n$ iterations. Reducible braids can be decomposed in distinct subbraids, where we can imagine ``tubes'' encompassing the subbraids, which then themselves form a braid. Finally, braids that are not finite-order or reducible are called pseudo-Anosov, and they represent well-entangled braids that cannot trivially be decomposed or untied. We will use the TN isotopy class to differentiate between twisted and braided polymer bundles.

\section{Aggregation and structural phase transitions \label{sec:simulation}}

To study the braiding properties of a multichain system, we simulated a coarse-grained bead-spring model system of $M=4$ chains, each with $N=30$ or $N=10$ monomers with mass $m$, using the Kremer-Grest model \cite{Grest1986,Kremer1990} by means of the LAMMPS software. The total number of monomers in the system is then $\mathcal{N} = M\cdot N$. The interaction potential between monomers is of Lennard-Jones (LJ) type, whose van der Waals length and cohesive strength scales are $\sigma$ and $\epsilon k_B T$, respectively. The hard-sphere monomer beads are connected by strong nonlinear springs, characterised by a finite-extensible-nonlinear spring (FENE) potential. The bending energy between two successive bonds spanning an angle  $\theta$ is equal to $U_{\rm bend}(\theta) = \kappa k_B T (1-\cos(\theta))$, where $\kappa$ is the bending parameter or elastic constant. In all simulations, we set $k_B = \epsilon = \sigma = 1$, such that the stiffness parameter and temperature tune the bending energy of the chains in units of $\epsilon$. The characteristic timescale is $\tau = \sigma\sqrt{m/k_B T}$ and the average bond length is $\ell_b = 0.965\sigma$ \cite{Hunt2009}. The time step for all simulations is set to $\Delta t = 0.01\tau$. After equilibration, the circuit topological content is recorded. For all configurations studied, we perform the simulations in the NVT ensemble, where we fix the temperature $T$ to be low enough as to be in the aggregation regime. Since we do not focus on the thermodynamics of the system in this work, we will not study thermodynamic phase transitions and will only vary the temperature as a means to control the aggregation of the system before varying the stiffness, ensuring that all polymers participate in the resulting amorphous or bundled aggregate.

\begin{figure}[htp]
    \centering
    \includegraphics[width=\linewidth]{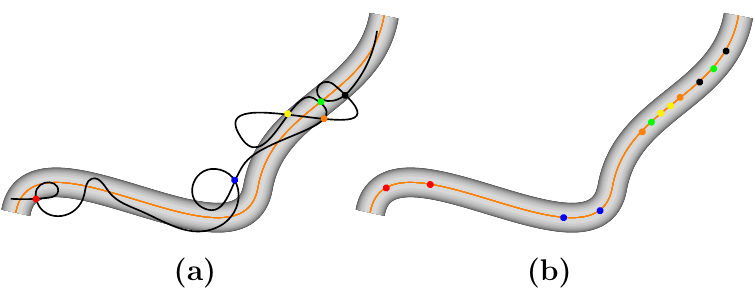}
    \caption{\textbf{(a)} The reduction of a fluctuating polymer (black line) within its confining tube (gray) to a primitive path (orange line). Hard contacts are indicated by colored points. \textbf{(b)} Topological arrangement of the individual hard contact points (colored) within the polymer's primitive path.}
    \label{fig:ppa}
\end{figure}

To determine the mutual entanglement of collections of fluctuating chains, a primitive path analysis (PPA) is subsequently carried out to reduce the system to a collection of straight segments that are interlocked \cite{Caraglio2017}. The PPA algorithm essentially contracts the contours of the chains while keeping the ends fixed and without allowing strand passages, see Fig.~\ref{fig:ppa}. In this manner, we combine circuit topology with entanglement and make subsequent braid analysis easier to perform.

After the PPA is performed, the spatial coordinates of the monomers are recorded and connected by means of a linear interpolation, which is justified due to the volume exclusion we use in the simulations and the PPA algorithm; no other monomer can occupy the space between two bonded monomers in the polymer backbone. The resulting ``trajectories'' are subsequently analysed by means of the Matlab package BraidLab \cite{braidlab,Thiffeault2022} and the writhe, braid length, complexity index and isotopy class are computed. More details on the computational aspects can be found in \cite{SM}.

We now study the braided circuit in two aspects: the circuit topological content and the braiding measures. For every configuration, the writhe, complexity, braid length and isotopy class are recorded. To track the structure of the polymer system, we will use the multichain radius of gyration $R_g$, which is defined as
\begin{equation}
    \label{eq:RG}
    R_g^2 = \frac{1}{\mathcal{M}}\sum\limits_{i=1}^{\mathcal{N}} m_i|r_i-r_{CM}|^2\,,
\end{equation}
where $\mathcal{M} = \sum_i m_i$ is the total mass of all monomers, $r_i$ and $m_i$ are respectively the position and mass of the $i$th monomer, and $r_{CM}$ is the center of mass of the full system. We subsequently make $R_g$ dimensionless by rescaling by the single-chain radius of gyration, i.e, $\widetilde{R}_g = R_g/R_{g,\theta}$, where $R_{g,\theta} = b\,\sqrt{(N-1)/6}$ is the entropically governed radius of gyration of a three-dimensional random walk with equilibrium bond length $b$ and stiffness $\kappa=0$.

Additionally, we define another end-to-end correlation parameter $C_R$ as 
\begin{equation}
    \label{eq:CR}
    C_R = \frac{2}{M (M-1)}\sum\limits_{i<j} \left({\bf \hat{R}_i}\cdot{\bf \hat{R}_j}\right)^2\,, 
\end{equation}
where ${\bf \hat{R}_i}$ is the unit end-to-end vector of the $i$th polymer \cite{Zierenberg_2015}. This parameter is $C_R =1$ for completely aligned bundles, and $C_R = 1/3$ for uncorrelated polymer systems, i.e., in the amorphous regime. It plays a role similar to a nematic order parameter.

Due to long equilibration times for longer polymers, we analyse the smaller $N=10$ system in more detail, averaged over a larger ensemble of configurations, while for the longer $N=30$ polymers we present more rough data.

\begin{figure*}[htp]
    \centering
    \includegraphics[width=\linewidth]{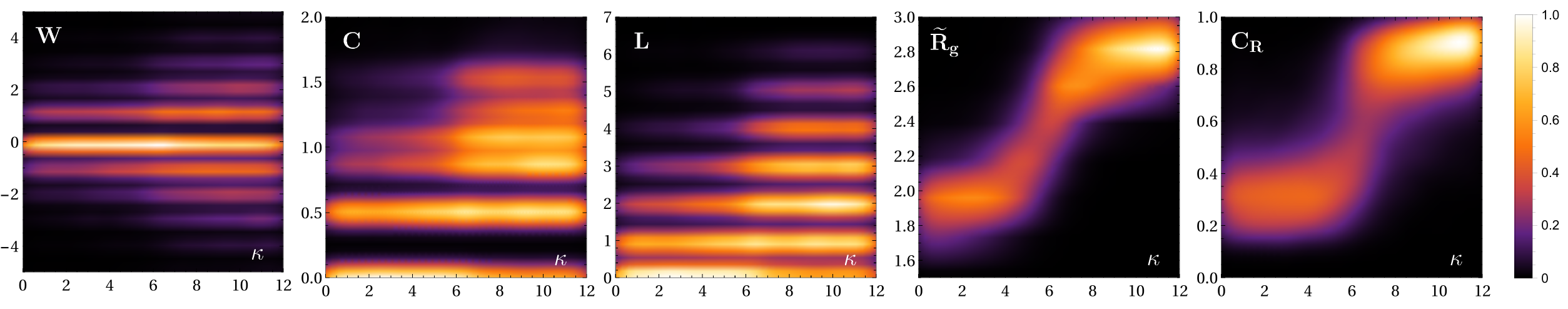}
    \caption{Density plots of the topological observables for $N=10$ as a function of the stiffness $\kappa$. It can be easily seen that there is a transition from an amorphous to an aligned aggregate around $\kappa \approx7$. The colours indicate the probability of an observable having a value given on the $y$-axis. For high stiffness values, all observables stabilise around a fixed distribution.}
    \label{fig:densitygridN10}
\end{figure*}

Let us first consider the case $N=10$ at a temperature of $T=1$ and density $\rho = 0.01$, where we vary the parameter $\kappa$ from $\kappa=0$ to $\kappa = 12$ in increments of $\Delta\kappa = 0.2$, covering the entire stiffness range. We average over 1000 realisations of the system per value of $\kappa$, after equilibrating the system for a time of $20000\tau$, such that $\widetilde{R}_g,\,C_R$ have saturated and only fluctuate around their steady-state value.

The end-to-end correlation parameter $C_R$ varies smoothly from $C_R \approx 1/3$ to $C_R \approx 0.84$, not completely reaching $C_R =1$. This indicates that the four chains are aligned along a common axis and that they have transitioned from an uncorrelated, amorphous state to a more ordered state, where the chains form an oriented bundle. These bundles can twist and form (braided) bundles in order to minimise their energy, or as a means for kink stabilisation as a result of defects \cite{Slepukhin2021}. This can be observed by studying the braid length and complexity, which are increasing functions of $\kappa$, and which saturate at the transition at $\kappa \approx 7$. Note also that the average writhe fluctuates around zero, as should be the case. A more detailed breakdown of the distributions of the aforementioned quantities as functions of $\kappa$ is shown in Fig.~\ref{fig:densitygridN10} 

The isotopy class also reveals information on the topology of the system. Generally, there is high fraction of finite-order braids, for all values of $\kappa$. A natural explanation for this behaviour is that since for low values of $\kappa$, the chains first collapse onto themselves and only then aggregate. As a result, when the PPA reveals the resulting braid $\beta$, it is unbraided, i.e., equal to the identity $\beta = e$. Since the latter is finite-order, we see an overabundance of this isotopy class for low values of $\kappa$. The increasing fraction of reducible and pseudo-Anosov braids for higher values of $\kappa$ is then a natural extension of this reasoning; when the chains become stiffer, they have a higher probability of aggregating and forming more complex braids. For the reducible isotopy class, a subset of the chains aggregate first and form separate braids, e.g., $\beta = \sigma_1\sigma_3$, and subsequently aggregate into a single structure. The pseudo-Anosov class then results from a collective aggregation and braiding. These statements about the isotopy class are not exhaustive; a pseudo-Anosov braid can as well result from taking a different projection of the system. As an example, consider the braid $\beta = \sigma_1\sigma_2$, in which three strands cross and the fourth is left unbraided. This is a reducible braid, with ``tubes'' around the first three, and around the fourth. By rotating our projection angle slightly, we could make the third strand undercross the fourth, making it a pseudo-Anosov braid. Since there is no preferred direction or external force that influences the system, the projection angle depends on the specifics of the polymer coordinates. We simulate the system multiple times, and thus we can assume that such artefacts can be neglected to have an influence. 

\begin{figure}[htp]
    \centering
    \includegraphics[width=\linewidth]{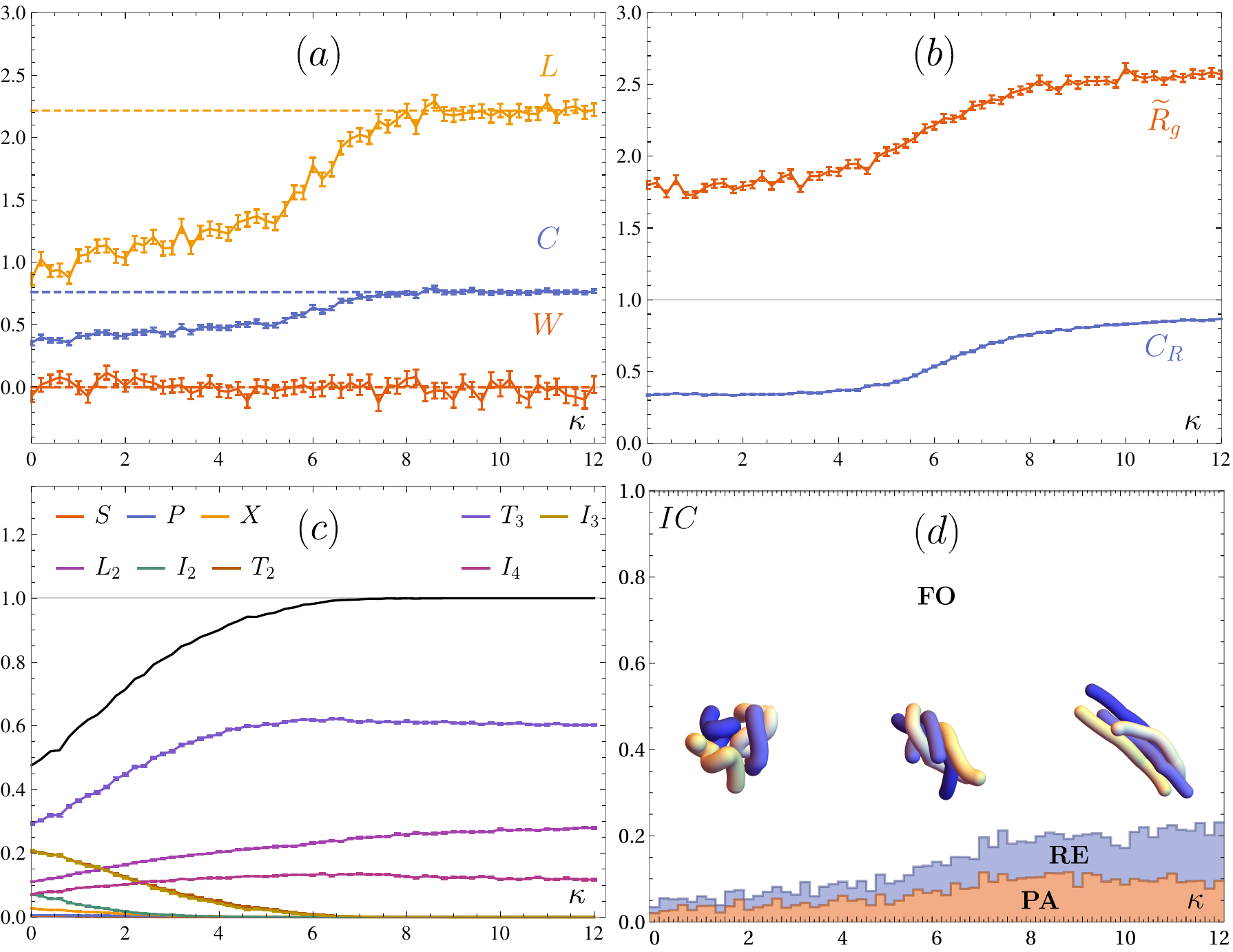}
    \caption{Observables in the polymer system with $M=4$, $N=10$ at $T=1$, $\rho=0.01$. {\bf (a)} Writhe $W$, complexity $C$ and length $L$. {\bf (b)} The rescaled radius of gyration $\widetilde{R}_g$ and correlation parameter $C_R$. {\bf (c)} Circuit topology fractions of the different motifs; error bars are smaller than symbol size. {\bf (d)} Isotopy class given by the TN types (RE, FO, PA), and representative structural conformations of the system at different $\kappa$. All results are averaged over 1000 runs.}
    \label{fig:N10gridplot}
\end{figure}

Let us now study the circuit topology in more detail. The ensemble averaged circuit topological motif fractions are shown in Fig.~\ref{fig:N10gridplot}(c); a more detailed breakdown of the distribution of the topological motif fractions as a function of the stiffness is given in the Supplemental Material \cite{SM}. From these figures, it can be seen that already for $\kappa=0$, the fraction of single-chain motifs, i.e., $S,\, P,\, X$ is very low. The reason is that these motifs require the formation of two loops on a single chain, which is highly unlikely for the short polymers we study here. As a result, the motifs that either involve two loops on two separate chains (i.e., $I_2$), or one loop on a single chain ($T_2$ and $I_3$), are somewhat more represented. Since in our setup we study the aggregated phase, the motifs that consist of only interchain connections, ($L_2$, $T_3$, $I_4$) are dominant for $\kappa \gtrsim 2$. Within the latter motifs, the relative abundance of $T_3$ is a consequence of the low number of polymers in the system; a disordered chain can more easily form two distinct bonds with two other chains, in contrast with forming two distinct bonds with the same chain, which tends to align both chains. Moreover, the $I_4$ motif requires that two bonds are distinctly \emph{not} sharing a polymer. Since the aggregates are closely packed, this situation is also more unlikely than either $T_3$ or $L_2$. Therefore, we see that, on average, $[T_3]>[L_2]>[I_4]$ for all values of $\kappa$. For increasing stiffness $\kappa$ the aggregate aligns and the motifs that contain loops vanish. It can then be seen that the motifs containing only interchain contacts, i.e, $L_2$, $T_3$ and $I_4$ tend to dominate the system. For $\kappa \gtrsim 7$, the sum of these motifs is equal to one, indicating that all intrachain contacts (i.e., loops) have disappeared. Hence, the total fraction of $L_2$, $T_3$ and $I_4$ motifs can be used as an order parameter to measure the degree of alignment of an aggregate.

\begin{figure}[htp]
    \centering
    \includegraphics[width=\linewidth]{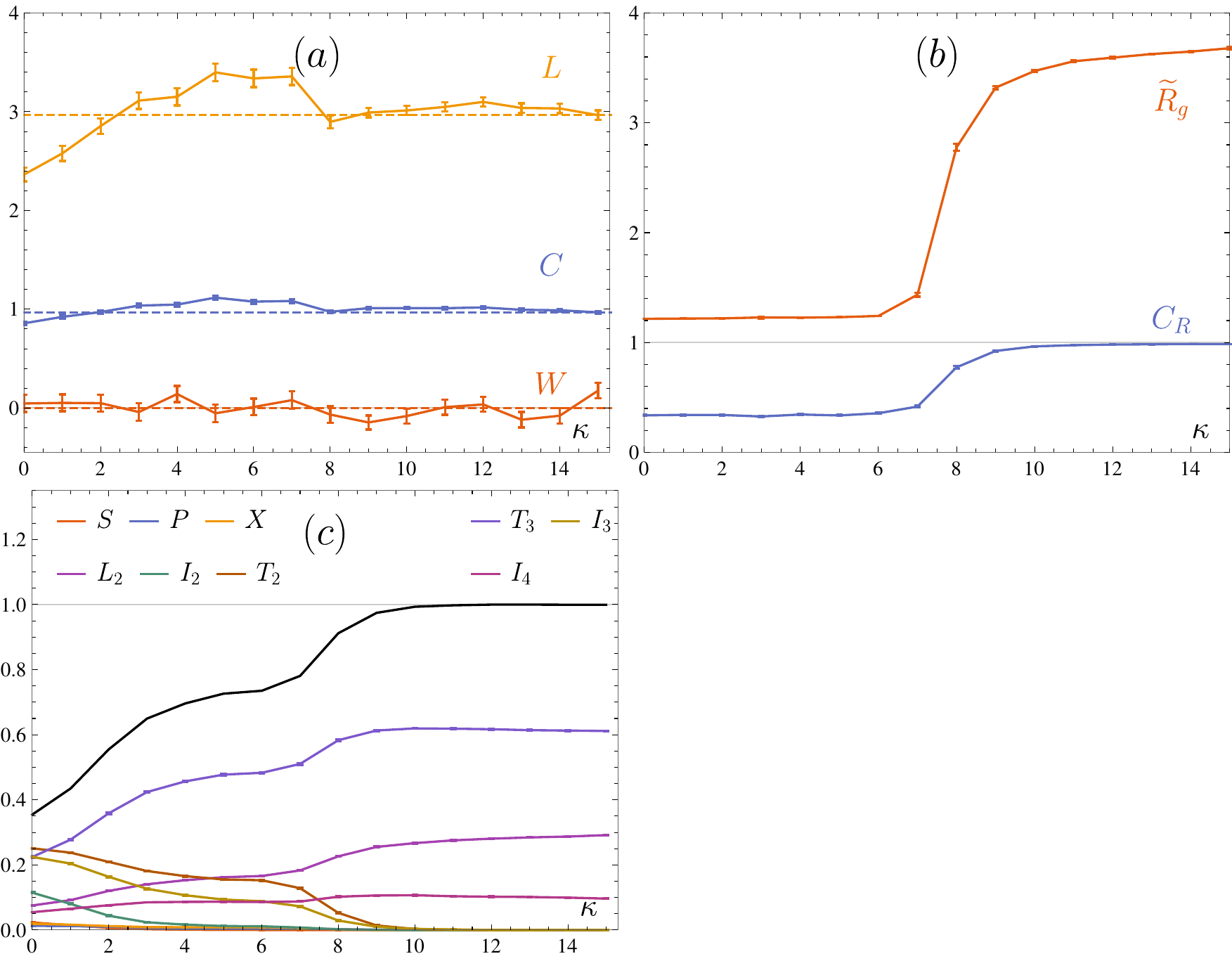}
    \caption{Observables in the polymer system with $M=4$, $N=30$ at $T=0.1$, $\rho=0.01$. {\bf (a)} Writhe $W$, complexity $C$ and length $L$. {\bf (b)} The rescaled radius of gyration $\widetilde{R}_g$ and correlation parameter $C_R$. {\bf (c)} Circuit topology fractions of the different motifs; error bars are smaller than symbol size. All results are averaged over 1000 runs.}
    \label{fig:N30gridplot}
\end{figure}

\begin{figure*}[htp]
    \centering
    \includegraphics[width=\linewidth]{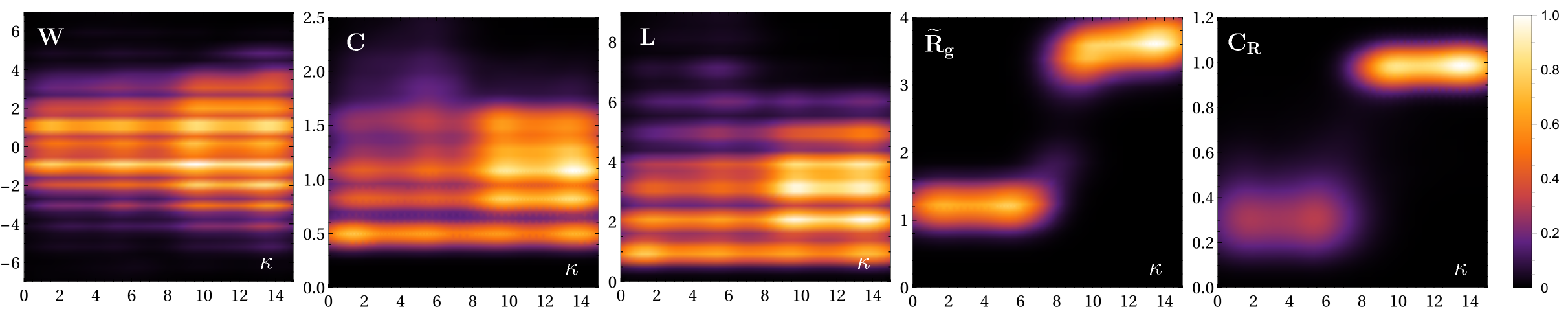}
    \caption{Density plots of the topological observables for $N=30$ as a function of the stiffness $\kappa$. It can be easily seen that there is a transition from an amorphous to an aligned aggregate around $\kappa \approx8$. The colours indicate the probability of an observable having a value given on the $y$-axis. For high stiffness values, all observables stabilise around a fixed distribution.}
    \label{fig:densitygridN30}
\end{figure*}

Let us now consider longer chains where the number of monomers is now $N=30$. At a temperature of $T=0.1$ and density $\rho=0.01$, we equilibrate the system for a time of $350000\tau$ until $\widetilde{R}_g$ has, on average, reached its equilibrium value. Taking stiffness steps of $\Delta \kappa =1$, we again compute the braiding quantities $W$, $C$ and $L$, along with $\widetilde{R}_g$, $C_R$ and the CT fractions. The results are shown in Figs.~\ref{fig:N30gridplot} and~\ref{fig:densitygridN30}.

In contrast with the $N=10$ system, we see that the topological quantities are not monotonous anymore, but exhibit a region of increased braid length and complexity for $\kappa\lesssim 8$, while for $\kappa\gtrsim8$ these quantities saturate at the values $L\approx3$ and $C\approx 1$. From this behaviour, it becomes clear that for very flexible chains individual collapse occurs before aggregation, decreasing the probability of finding heavily entangled aggregates. When the stiffness increases, however, there is a competition between aggregation and collapse; the individual chains can now aggregate and subsequently collectively collapse to form more intricate entangled structures. As a consequence, the length and complexity increase while $\widetilde{R}_g$ and $C_R$ indicate that the system is still not aligned and are unable to capture this behaviour. For a stiffness of $\kappa\approx8$, the individual chains aggregate but do not collapse, leading to lower average values for the braid length and complexity. 

Note that $L$ and $C$ converge to the values $L=3$ and $C =1$ or, equivalently, the ratio $C/L$ converges to $1/3$, which corresponds to the braid $\beta = \sigma_1^{\pm1}\sigma_2^{\pm1}\sigma_3^{\pm 1}$ (and isotopic equivalent braids). For four strands, this indicates that a single strand crosses over (under) all others; none of the strands in such a braid are interwoven. Consequently, the aligned polymer bundle is not twisted nor braided, since this situation is a result of the choice of projection angle. We can see that this also holds for the $N=10$ system, where $C/L \approx 0.346$.

Considering now the CT fractions, shown in Fig.~\ref{fig:N30gridplot}(c) and in the Supplemental Material \cite{SM}, we see a distinct difference with respect to the $N=10$ system. While the overall monotonous behaviour of the fractions remains the same as before, a plateau arises for values $4\lesssim \kappa \lesssim 7$, when the braiding quantities reach their maximal value. In this region, one can see that there is coexistence between $T_2$ and $L_2$, and between $I_3$ and $I_4$ motifs. This is a consequence of the energy balance between the formation of intra- and interchain contacts, i.e., forming loops versus forming interchain bonds. Since loops can enclose or trap other chains, they generally lead to a higher degree of braid entanglement. Hence, in the coexistence region, the braiding measures saturate at their maximal value as a consequence of loops being broken.

For larger stiffness, aggregation wins out over the collapse and the CT fractions associated with interchain interactions, i.e., $L_2$, $T_3$ and $I_4$ again dominate the motif space, since the polymers are stretched along a single direction when forming the aggregate. While $\widetilde{R}_g$ and $C_R$ might be able to detect the structural phase transition between the amorphous and aligned configurations, circuit topology provides a more nuanced and rich approach to the topology of entangled systems.

Considering the $N=30$ system, we can consider the absence of the plateau phase for the $N=10$ situation. It is indeed possible that such a plateau phase exists for the latter situation, but since the chains are too short to form complex braided aggregates, the stiffness range for which this occurs is too small to see in simulations. 

\section{Discussion and conclusions \label{sec:conclusion}}
In summary, our investigation of aggregates of semiflexible polymers has provided insights into their structural properties and phase transitions. We observed that varying the polymer stiffness leads to a structural "phase diagram" where different conformations such as amorphous or extended structures are favoured, depending on the stiffness. By adjusting the latter, we were able to study the competition between collapse and stiffening effects. The inter- and intrachain contacts provided valuable information about the system's structural properties and phase transitions, which could be analyzed within the framework of braided multichain circuit topology (CT). The combination of circuit topology and entanglement analysis allowed for a more comprehensive understanding of aggregation processes in systems with a small number of chains.

The analysis of braided circuits revealed the importance of topological measures such as writhe, complexity, braid length, and isotopy class. The isotopy class analysis showed a higher fraction of finite-order braids for lower stiffness values, while higher stiffness values favoured more complex pseudo-Anosov and reducible braids. The circuit topological analysis reveals the dominance of interchain connections, particularly motifs $T_3$, $L_2$, and $I_4$, which reflects the closely packed nature of the aggregates.

When considering longer chains, we observed a region of increased braid length and complexity, indicating a competition between aggregation and collapse. For very flexible chains, individual collapse occurred before aggregation, reducing the probability of finding heavily entangled aggregates. However, as stiffness increased, the chains aggregated and collectively collapsed, forming intricate entangled structures. The length and complexity of the braids increased in this regime, while the radius of gyration and end-to-end correlation parameter indicated that the system was not fully aligned. The circuit topology fractions revealed differences compared to the smaller system, with a plateau phase emerging for intermediate stiffness values when the braiding quantities reached their maximum. For larger stiffness, the dominance of interchain motifs indicated the stretching of polymers along a single direction during aggregate formation.

Our study provided a comprehensive understanding of the structural properties, phase transitions, and braiding characteristics of multichain systems. The combination of circuit topology analysis and braiding measures allowed for a deeper exploration of the system's behavior, shedding light on the interplay between collapse, aggregation, and entanglement. These findings contribute to the broader understanding of semiflexible polymer systems and pave the way for further investigations in this field. One future application we envision is the study of the coil-globule transition for water-soluble hydrophobic polymer chains in aqueous solutions \cite{Hatano2016}. In such systems, the explicit presence of a polar solvent weakens the attractive interactions among monomers at temperatures close to room temperature and the polymers collapse upon heating, which is in contrast to polymers in inorganic solvents. Circuit topology can then be used to study the conformational changes of such systems at the critical temperature. Finally, the general framework of braided circuit topology is expected to be applicable to the broader field of materials science, including the design and analysis of nanotube, nanofiber, and nanowire assemblies. 

\section{Limitations of the study \label{sec:Limitations}}
Since we only study short identical polymers ($N=10$ or $N=30$ monomers), the current results might not be directly applicable to longer chains or to systems with a non-uniform length distribution; crossings can occur more often when the chain length increases and endpoint defects can induce a higher degree of braiding. Moreover, we here considered systems of only $M=4$ chains, while it is known that $M$ strongly influences the structural collapse and aggregation transitions \cite{Zierenberg_2015}. For systems where $M\geq4$, we expect the CT motif fractions to heavily favour the $I_4$ motif when stiffness increases. These limitations, however, can be readily addressed in future research with the same methodology and minor software adjustments.





\begin{acknowledgments}
We are grateful for inspiring discussions with I. Diamantis on the topic of braids, and to K. Koga for suggesting the water-soluble polymer chains as an avenue for future research.
\end{acknowledgments}

\subsection{Author contributions}
{\bf J.B.}: Conceptualization, Formal Analysis, Methodology, Software, Writing- Original draft preparation, Visualization.\\
{\bf A.M.}: Conceptualization, Supervision, Methodology, Writing- Original draft preparation, Project administration, Funding acquisition, Resources.

\subsection{Declaration of interests}
The authors declare no competing interests.

\bibliography{z_biblio.bib}

\end{document}


\title{Aggregation and structural phase transitions of semiflexible polymer bundles: a braided circuit topology approach
\\ -- Supplementary Material --}

\author{Jonas Berx}
\affiliation{Medical Systems Biophysics and Bioengineering, Leiden Academic Centre for Drug Research, Leiden University, 2333CC Leiden, The Netherlands}
\author{Alireza Mashaghi}
\email{a.mashaghi.tabari@lacdr.leidenuniv.nl}
\affiliation{Medical Systems Biophysics and Bioengineering, Leiden Academic Centre for Drug Research, Leiden University, 2333CC Leiden, The Netherlands}

\date{\today}
\pagenumbering{gobble}


\section{Qualitative description of the braid analysis algorithm}\label{app:A}

Here we provide computational details on the calculation of the different braiding properties and the circuit topology fractions. The subsection ordering indicates the workflow of the different algorithms.

\subsection{Circuit topology fractions}

The first step in the simulation workflow is the LAMMPS simulation of the system of polymers. Since a detailed description of this simulation step is given in the main text, we no not repeat it here. The output of the LAMMPS simulation consists of a trajectory file that indicates the coordinates of all monomers in the simulation box. The circuit topology algorithm calculates the mutual Euclidean distance between pairs of monomers and builds a list of contacts based on a user-given cutoff distance $r_c$. In our simulations, we choose this cutoff distance to be $r_{c} = 1.4\sigma$. For the self-interaction of monomers on the same chain, we also choose an interaction cutoff at $r_{c,s} = 3\sigma$, such that next-nearest neighbours on the same polymer backbone cannot constitute a contact, preventing the polymer from forming loops that only consist of two bonds. For increasing stiffness, the bending rigidity prevents the formation of such contacts anyway, so this choice is only of importance for low $\kappa$.

In a next step, the mutual relation between pairs of contacts, i.e., four distinct contact sites we number $i = 1,\,2,\,3,\,4$, is determined based on the following conditions:
\begin{enumerate}
    \item If two of the contact sites that correspond to one contact (for simplicity choose sites 1 and 2) are on the same polymer $p_1$, then only motifs $S,\, P,\, X,\, T_2,\, I_2,\, I_3$ , i.e., motifs with at least one loop, are possible. We then check the mutual relation of the other sites 3 and 4:
    \begin{enumerate}
        \item If the remaining sites also share the same polymer $p_2$, we can either have:
        \begin{enumerate}
            \item If {\bf $p_2 = p_1$}: the motif is one of the single-chain motifs $S,\, P\, X$. The exact motif can be determined by considering the string notation of the contacts.
            \item If $p_2\neq p_1$: the only possibility is the independent $I_2$ motif.
        \end{enumerate}
        \item The remaining sites are on different polymers. The only combinations can be $T_2$ or $I_3$. To proceed, we check on which polymer the contacts 3 and 4 are located. Name these respectively $p_3$ and $p_4$:
        \begin{enumerate}
            \item If $p_3 = p_1$ or $p_4 = p_1$: the polymer $p_1$ possesses three contact sites, so the resulting motif is $T_2$.
            \item $p_3\neq p_1$ and $p_3\neq p_1$: the sites are on different polymers, so we have three participating polymers. The resulting motif is $I_3$.
        \end{enumerate}
    \end{enumerate}
    \item If two of the contact sites that correspond to one contact (for simplicity choose sites 1 and 2) are on different polymers $p_1$ and $p_2$, we can have one of the following motifs: $L_2,\,T_2,\,T_3,\,I_3,\,I_4$. We then check the mutual relation of the other sites 3 and 4:
    \begin{enumerate}
        \item If sites 3 and 4 share the same polymer $p_3$, we can only have $T_2$ or $I_3$. 
        \begin{enumerate}
            \item If $p_3 = p_1$ or $p_3 = p_2$: the polymer $p_3$ possesses three contact sites, so the resulting motif is $T_2$.
            \item $p_3\neq p_1$ and $p_3\neq p_2$: the sites are on different polymers, so we have three participating polymers. The resulting motif is $I_3$.
        \end{enumerate}
        \item If sites 3 and 4 are on different polymers $p_3$ and $p_4$, we can have either of the following motifs: $L_2,\, T_3,\,I4$. We make the following distinction:
        \begin{enumerate}
            \item If none of the contact sites are on the same polymer, we have motif $I_4$
            \item If either one of the following conditions is fulfilled, we have motif $T_3$:
            \begin{itemize}
                \item $p_3 = p_1$ and $p_4\neq p_2$
                \item $p_3 = p_2$ and $p_4\neq p_1$
                \item $p_4 = p_1$ and $p_3\neq p_2$
                \item $p_4 = p_2$ and $p_3\neq p_1$
            \end{itemize}
            \item If none of the antecedent conditions are fulfilled, we have the $L_2$ motif.
        \end{enumerate}
    \end{enumerate}
\end{enumerate}

The absolute number of motifs then exported to a text file for further processing.

\subsection{Primitive path analysis}
The PPA is performed in LAMMPS by means of the PPA package \cite{Sukumaran2005,HAGITA2021123493}. The terminal points of the chains are fixed in space, and all interactions within each chain are disabled, with the exception of the polymer backbone FENE bonds. However, interchain excluded volume interactions in the system are retained, and subsequently the system's energy is reduced through minimisation. As a result, the chains eventually transform into a configuration composed of contiguous straight segments linked by distinct, relatively sharp angles. 

The primitive path analysis allows for a correct representation of the entanglement of a multichain system and provides the starting point for the braid analysis. The coordinates of the monomers are exported to a text file.

\subsection{Braiding topology analysis}
The text file generated by the PPA is imported into a Matlab script. The braid analysis is performed by means of the BraidLab package \cite{braidlab,Thiffeault2022}. This package requires the input to consist of $x$, $y$ and $z$ coordinates for different trajectories, or in this case, polymers. The $z$ coordinate, however, needs to be common for all polymers (the analogy for the trajectories is the equivalence of $z$ to a global time coordinate for trajectory sampling). Since the monomers do not share the same $z$ coordinates, the system first needs to be reshaped.

For simplicity, let us henceforth label the endpoints of polymer $i$ by $i$ and $i'$. To reshape, the algorithm picks one of the endpoints (for simplicity, endpoint 1), and calculates the distance to all other polymer endpoints, excluding the $1'$ endpoint. The closest endpoint is labelled as endpoint 2 and the vector ${\bf r}_{12}$ is calculated. The distance from this vector to the remaining endpoints, excluding $1'$ and $2'$ is calculated and the closest endpoint is labelled as $3$. Together, points $1,\,2,\,3$ span a plane $\Pi$, for which the normal vector ${\bf n}$ is calculated. For the remaining endpoints $\{4,\,4',\,1',\,2',\,3'\}$, the Euclidean distance $d$ between every endpoint and its orthogonal projection on the plane $\Pi$ is calculated. For the maximal distance $d_{\rm max}$, a plane $\Pi'$ that is parallel to $\Pi$ and passes through the endpoint corresponding to $d_{\rm max}$ is calculated.

In the next step, all remaining endpoints are projected onto the closest plane $\Pi$ or $\Pi'$ and the projections are relabelled as the new endpoints $i$ or $i'$, such that $i\subset\Pi$ and $i'\subset\Pi'$. 

The normal vector ${\rm n}$ between the parallel planes now constitutes the new $z'$-axis, the initial vector ${\bf r}_{12}$ we choose to be the new $x'$-axis after normalisation. A suitable third orthonormal vector can be found by using Gram-schmidt orthogonalisation. The old coordinates $\bf r$ can then be transormed to the new coordinate system $\bf r'$. The monomer coordinates, as well as possible new endpoints $i,\,i'$ are subsequently connected by means of a linear interpolation, which is justified due to the volume exclusion we use in the simulations and the PPA algorithm; no other monomer can occupy the space between two bonded monomers in the polymer backbone. The resulting system now possesses a common $z'-$axis and the polymer 'trajectories' can be partitioned into two-dimensional 'slices' with equal $z'$ coordinate.

These sliced trajectories are subsequently used as input for the BraidLab package and the writhe, complexity, minimal braid length and isotopy class are recorded for every configuration in the simulated ensemble.

\section{Distributions of the circuit topology motif fractions}
\begin{figure*}[htp]
    \centering
    \includegraphics[width=\linewidth]{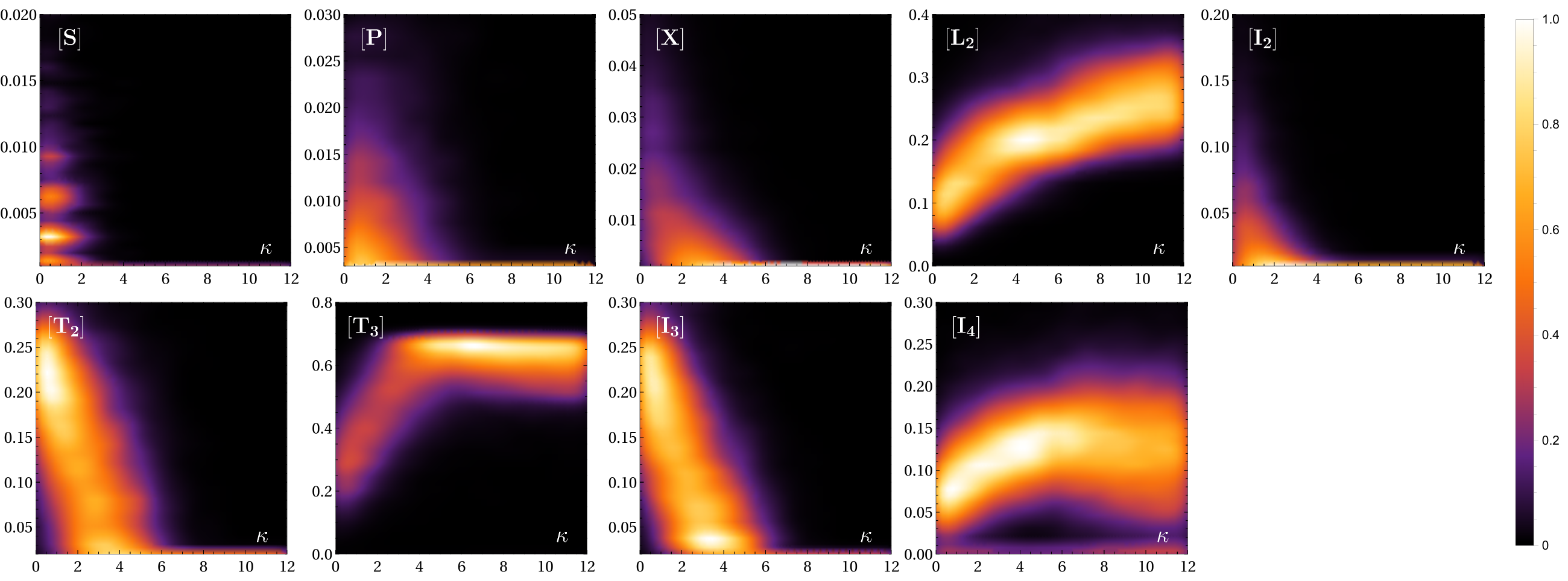}
    \caption{Density plots for the topology fractions of the different CT motifs as a function of stiffness in the system with $N=10$ monomers per chain.}
    \label{fig:CTgridN10}
\end{figure*}

\begin{figure*}[htp]
    \centering
    \includegraphics[width=\linewidth]{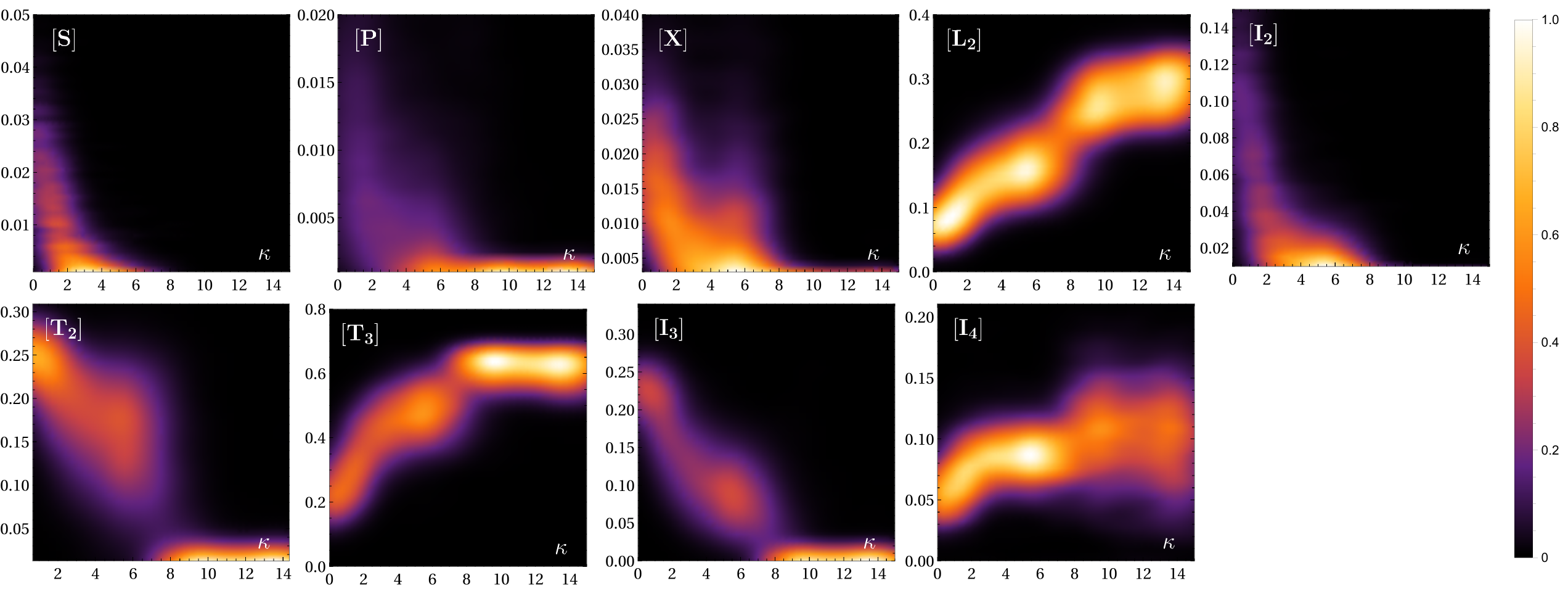}
     \caption{Density plots for the topology fractions of the different CT motifs as a function of stiffness in the system with $N=30$ monomers per chain.}
    \label{fig:CTgridN30}
\end{figure*}

We provide here a detailed breakdown of the distribution of the topological motif fractions as a function of the stiffness $\kappa$ for $M=4$ and $N=10$ (Fig. \ref{fig:CTgridN10}) or $N=30$ (\ref{fig:CTgridN30}). The colours indicate the probability of a motif having a given fraction on the $y$-axis. For high stiffness values, all motif fractions stabilise around a fixed distribution.

\bibliography{z_biblio.bib}